\documentclass[aps,pre,reprint,groupedaddress,showpacs]{revtex4-1}

\usepackage{graphicx}
\usepackage{url}

\begin{document}

\title{Fractional Brownian motion and the critical dynamics of zipping polymers}

\author{J.-C. Walter}
\affiliation{Institute for Theoretical Physics, KULeuven, Celestijnenlaan 200D, B-3001 Leuven, Belgium}
\author{A. Ferrantini}
\affiliation{Institute for Theoretical Physics, KULeuven, Celestijnenlaan 200D, B-3001 Leuven, Belgium}
\author{E. Carlon}
\affiliation{Institute for Theoretical Physics, KULeuven, Celestijnenlaan 200D, B-3001 Leuven, Belgium}
\author{C. Vanderzande}
\affiliation{Faculty of Sciences, Hasselt University, Agoralaan 1, B-3590 Diepenbeek, Belgium}
\affiliation{Institute for Theoretical Physics, KULeuven, Celestijnenlaan 200D, B-3001 Leuven, Belgium}

\begin{abstract}
We consider two complementary polymer strands of length $L$ attached
by a common end monomer. The two strands bind through complementary
monomers and at low temperatures form a double stranded conformation
(zipping), while at high temperature they dissociate (unzipping).
This is a simple model of DNA (or RNA) hairpin formation.  Here we
investigate the dynamics of the strands at the equilibrium critical
temperature $T=T_c$ using Monte Carlo Rouse dynamics. We find that the
dynamics is anomalous, with a characteristic time scaling as $\tau \sim
L^{2.26(2)}$, exceeding the Rouse time $\sim L^{2.18}$.  We investigate
the probability distribution function, the velocity autocorrelation
function, the survival probability and boundary behavior of the underlying
stochastic process. These quantities scale as expected from a fractional
Brownian motion with a Hurst exponent $H=0.44(1)$.  We discuss similarities
and differences with unbiased polymer translocation.
\end{abstract}

\date{\today}


\maketitle


\def \bc{\begin{center}}
\def \ec{\end{center}}
\def \beq{\begin{equation}}
\def \eeq{\end{equation}}
\def \beqa{\begin{eqnarray}}
\def \eeqa{\end{eqnarray}}
\def \widwide{13cm}
\def \widnormal{8cm}
\def \widsmall{5cm}
\def \kB{{\mathrm{k_B}}}
\def \kb{{\mathrm{k_B}}}
\def \Lk{{\mathit{Lk}}}
\def \Tw{{\mathit{Tw}}}
\def \Wr{{\mathit{Wr}}}

\renewcommand{\thefootnote}{\fnsymbol{footnote}}

\def \widAA{12cm}
\def \widA{8cm}
\def \widAb{7cm}
\def \widB{6.1cm}
\def \widBB{3.cm}
\def \widBBB{2.cm}

\def \uu{\hat{u}}
\def \uuab{{u_{\rm AB}}}
\def \vv{\hat{v}}
\def \oo{\infty}
\def \fpG{$\mathcal{G}$}
\def \fpD{${\mathcal{S}}_0$}
\def \fpGb{${\mathcal{G}}_{0}$}
\def \fpS{$\mathcal{S}$}
\def \bb{(\beta-\beta_c)}
\def \btt{\begin{tt}}
\def \ett{\end{tt}}

\newcommand{\chapterC}[1]{   \chapter{#1}  }
\newcommand{\chapterCx}[1]{  \chapter*{#1}  }
\newcommand{\sectionC}[1]{   \section{#1}  }
\newcommand{\sectionCx}[1]{   \section*{#1}  }
\newcommand{\captionC}[1]{\caption{\footnotesize{\sf #1}}}

\section{Introduction}

There has been an ongoing interest in recent years in the analysis of
models of polymer dynamics. The origin of this interest is due to two
main facts. Firstly, experiments allow nowadays to control polymers at
nanoscales and to follow the behavior of single molecules, providing
thus many insights on their dynamics \cite{lumm03,shus04,petr06}. This
has motivated more theoretical research in the field. Secondly, many of
these systems show an anomalous dynamics, a paradigmatic example being
that of a polymer translocating through a nanopore \cite{chua01}. Modeling
anomalous dynamics has attracted quite some attention in the Statistical
Physics community due to the ubiquity of this behavior in many physical
systems as disordered media~\cite{bouc90}, conformational fluctuations
of proteins~\cite{kou04}, diffusion of molecules in cells \cite{jaeh11}
and polymers \cite{panj10}.

In the case of polymer translocation, the subdiffusive behavior
is inferred from the scaling $\tau \sim L^\alpha$ with $\alpha >
2$, which relates the translocation time $\tau$ to the polymer
length $L$.  Although there have been a large number of publications
\cite{sung96,muth99,chua01,wolt06} there is no general agreement
on the value of $\alpha$ obtained from simulations. Also on the
theoretical side different predictions for $\alpha$ have been
made~\cite{chua01,panj07,saka10}.

\begin{figure}[t]
\centering\includegraphics[width=0.48\textwidth]{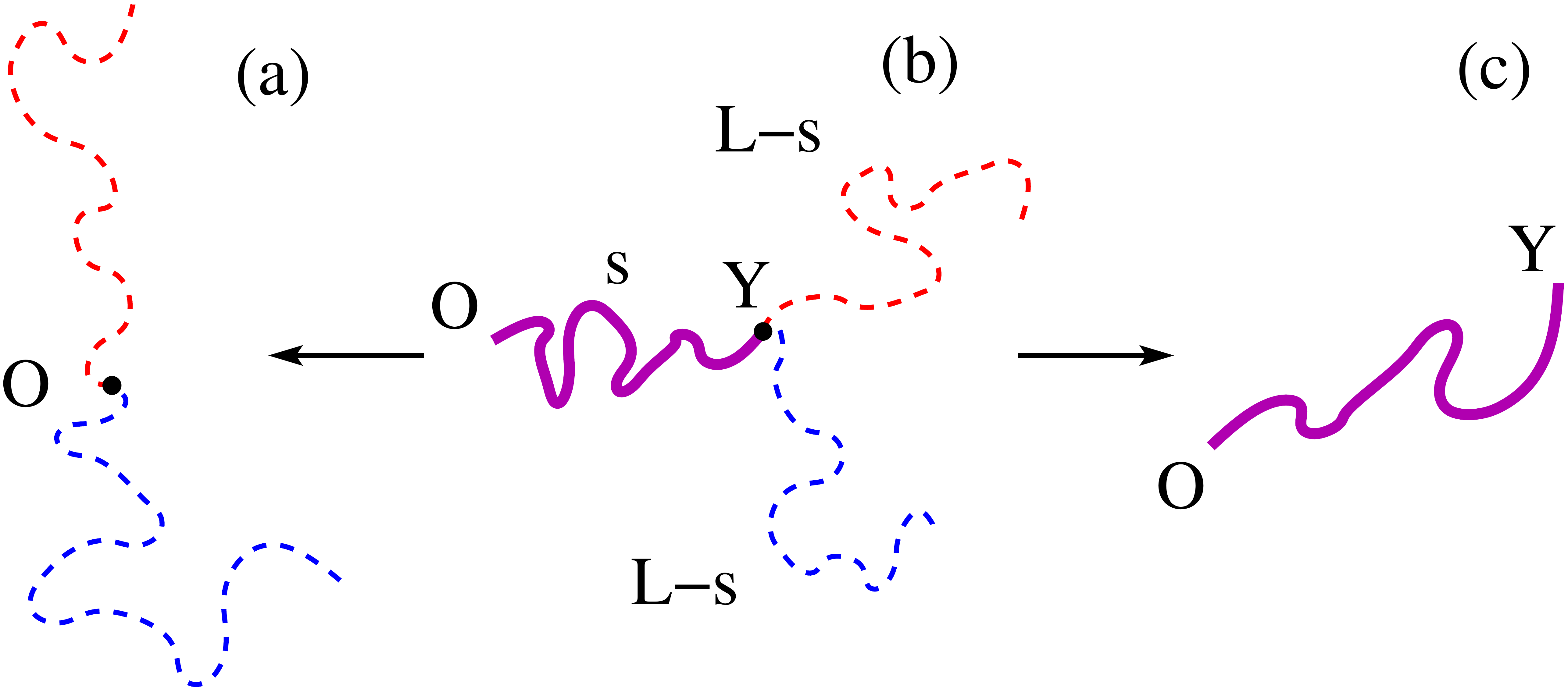}
\caption{(Color online) Sketch of the zipping/unzipping dynamics. The two
polymer strands are joined by a common end O. Single stranded polymers are
shown as dashed lines, while double stranded as a thick solid line.  (a)
Fully unzipped state, (b) Partially zipped, (c) Fully zipped.  Y denotes
the position of the end of the double stranded segment.  The total length
of each strand is $L$. A partially zipped configuration is characterized
by two single strands of length $L-s$ and a double strand of length $s$.}
\label{zipunzip}
\end{figure}

Besides translocation, there are other polymer processes which are
expected to show anomalous dynamics.  We are interested here in the
analysis of zipping dynamics, which is the process through which two
``complementary" strands attached by one end bind/unbind from each other,
as shown in Fig.~\ref{zipunzip}.  In our model we do not allow bubbles
to be formed, hence the dynamics proceeds sequentially as in a zipper.
We focus here on the dynamics at the transition temperature $T=T_c$.
At low temperatures ($T < T_c$) the system is driven towards the fully
zipped state (Fig.~\ref{zipunzip}(c)), whereas at high temperatures ($T >
T_c$) the two strands unbind (Fig.~\ref{zipunzip}(a)). The dynamics for
these two cases were studied in Ref.~\cite{ferr11}. Interestingly, the
zipping time was found to scale as the length of the polymer as $\tau_{z}
\sim L^\alpha$ with $\alpha \approx 1.37$, whereas the unzipping time
scaled as $\tau_{u} \sim L$.  If the dynamics were sufficiently slow (see
below), so that zipping and unzipping proceed through quasi-equilibrium
states, the times would scale linearly in the strand length $L$. This
corresponds to the motion of a Brownian particle, the fork point Y shown
in Fig.~\ref{zipunzip}(b), in a linear downhill potential. A scaling
$\tau_{z} \sim L^\alpha$ with $\alpha \neq 1$ implies anomalous dynamics.
We note, in addition, that the anomalous exponent for zipping ($\alpha
=1.37$) turns out to be in agreement with that found in some simulations
of forced polymer translocation \cite{vock08,luo09}.

We recall that in forced (or biased) translocation an external field
drives the polymer preferentially towards one of the two sides of the
separating membrane. This can be realized experimentally, for instance
for a charged polymer as DNA or RNA by imposing an electric potential
difference on the two sides. In unbiased translocation the external field
is absent and the polymer translocation is driven by thermal fluctuations.

The aim of this paper is to extend the analysis of the zipping
dynamics to the critical temperature $T=T_c$. In this case there
is no strong bias towards either the zipped or unzipped state, as
these two have the same free energy in the thermodynamic limit. One
could therefore expect an analogous behavior of that found in unbiased
polymer translocation. However, the exponent we find in our simulations
disagrees with those conjectured for unbiased translocation suggesting
that critical zipping is in another universality class.  We find however
that the underlying stochastic process is well-described by a fractional
Brownian motion \cite{mand68} for which we determine the Hurst exponent,
probability distribution function and survival probabilities.

\section{Model}

The model discussed here was also used in recent studies of renaturation
dynamics \cite{ferr10} and zipping dynamics \cite{ferr11}.  Two polymers
defined on a face-centered-cubic (fcc) lattice are joined by a common
end.  The monomers on both strands are labeled by an index $i=0,1,
\ldots , L$. The two strands are self- and mutually avoiding, with
the exception of monomers with the same index $i$, which are referred
to as complementary monomers.  Two complementary monomers can bind by
overlapping on the same lattice site.  A typical simulation run start
with all monomers $0 \leq i \leq L/2$ bound, and the monomers $i > L/2$
unbound (see Fig.~\ref{zipunzip}(b)).  This initial configuration is
relaxed to equilibrium by means of pivot \cite{madr88} and local moves
which leave the number of bounded monomers unchanged.  Once an equilibrium
configuration is obtained the actual simulation is started. The pivot
algorithm is no longer used and the Monte Carlo updates are strictly
local. They consist of corner-flip or end-flip moves that do not violate
self- and mutual avoidance. 
{\bf A feature of the fcc lattice geometry is that the local moves are
such that two different polymer chains cannot intersect each other. In
the simple cubic geometry the non-crossing condition can be realized by
the bond-fluctuation model \cite{carm88}. The local dynamics on the fcc
lattice reproduces the Rouse model behavior, as shown in~\cite{thesis}.
}

A Boltzmann weight $\omega > 1$ is associated to the binding of two
monomers. In the Monte Carlo dynamics binding occurs with probability $1$
while unbinding with probability $1/\omega$, so that detailed balance
is satisfied.  A Monte Carlo step consists in selecting a random monomer
on one of the two strands. If the selected monomer is unbound a local
flip move is attempted. If the selected monomer is a bound monomer there
are two possibilities. Either a local flip of the chosen monomer is
attempted, and if accepted, this move results in the bond breakage; or a
flip move of both bound monomers is generated, which does not break the
bond between them.  In the model discussed here we do not allow any bubble
formation neither for zipping nor unzipping, by imposing the constraint
that monomer $i$ can bind to its complement only if monomer $i-1$
is already bound. Analogously monomer $i$ can unbind only if monomers
$i+1$ are already unbound. This is the model Y which was referred to in
Ref.~\cite{ferr11}.  As unit of time we take $2L+1$ Monte Carlo steps,
so that in one time step one attempted update per monomer is performed.

\section{Equilibrium free energies}

We start by discussing some properties of the equilibrium free energy of
zipping polymers. We consider in particular the dependence of this free
energy on the coordinate $s$, describing the position of a fictitious
Brownian particle. We analyze the motion of this particle in the given
free energy landscape.

The number of configurations for a linear polymer of length $L$,
in the limit of $L \to \infty$, takes the asymptotic form $Z \sim
\mu^L L^{2 \sigma_1}$ where $\mu$ is the connectivity constant and
$\sigma_1$ a universal exponent associated to the end vertex of
the polymer~\cite{vand98} (we use here a different notation from the
customary exponent $\gamma = 1 + 2 \sigma_1$).  In general the partition
function of polymer networks of more complex topology~\cite{dupl86}
is also characterized by subleading universal exponents.  For instance
a star polymer with three arms of length $L$ has a partition function
scaling as $Z_3 \sim \mu^{3L} L^{\sigma_3 + 3 \sigma_1}$, with $\sigma_3$
the exponent associated to a vertex with three outgoing arms. Here the
factor $3 \sigma_1$ accounts for the three end vertices.

We can now consider the case of partially zipped polymer strands of
Fig.~\ref{zipunzip}(b), which is related to that of a three arms star
polymer. We allow now the three arms to have lengths $L-s$ and $s$.
In addition we have to account for the Boltzmann weight $\omega$
associated to the binding between the $s$ monomers on the double
stranded segment. The total partition function is then given by
\begin{equation}
Z(s) = \mu^{2L-s} \omega^s L^{3 \sigma_1 +\sigma_3} f(s/L)
\label{z_3star}
\end{equation}
where $f(x)$ a scaling function. We analyze now the limiting scaling
behavior of the partition function (\ref{z_3star}).  For $s \to 0$
one should recover the partition function of a single polymer of length
$2L$ which yields $f(x) \sim x^{\sigma_1+\sigma_3}$. The analysis of the
limit $s \to L$ similarly imposes $f(x) \sim (1-x)^{\sigma_1+\sigma_3}$.
One can incorporate the two limits in the following expression:
\begin{equation}
Z(s) = \mu^{2L-s} \omega^s L^{3 \sigma_1 +\sigma_3} 
\left( \frac s L \right)^{\sigma_1+\sigma_3}
\left( 1 - \frac s L \right)^{\sigma_1+\sigma_3}
g(s/L)
\end{equation}
with $g(x)$ an analytic function of its variable.

We distinguish now two cases: $\omega \neq \mu $ and $\omega = \mu$, which
correspond to the off-critical and to the critical case, respectively.
If $\omega \neq \mu$ the leading dependence on $s$ of the free energy is
\begin{equation}
f(s) = - \log Z(s) \sim -s \log(\omega/\mu)
\label{free_en_offTc}
\end{equation}
where the constant terms, and subleading terms in $s$ have been omitted.
For $\omega > \mu$ the zipped state (low temperature phase) is favored,
whereas $\omega < \mu$ favors unzipping (high temperatures). Viewing
$s$ as a zipping coordinate one can consider a Fokker-Planck (FP)
description of the process, which is the motion of a Brownian particle
on a potential $f(s)$. For $\omega \neq \mu$ the potential is linear in
$s$ (Eq.~(\ref{free_en_offTc})), equivalent to a biased Brownian motion
towards $s =0$ ($\omega < \mu$) or $s=L$ ($\omega > \mu$). In both cases
the characteristic time of the particle for starting in one boundary to
reach the opposite one scales as $\sim L$.

At the critical point ($\omega = \mu$) the free energy becomes
\begin{equation}
f(s) = -(\sigma_1 + \sigma_3) \log\left[ \frac s L  
\left(1 - \frac s L\right) \right]
- \log g(s/L)
\label{free_tc}
\end{equation}
(omitting terms which do not depend on $s$).  We compare now this
expression with the free energy of translocating polymers. Consider
a polymer on a pore of an infinitely wide separating plane. The pore
divides the polymer into two non interacting parts of lengths $s$
and $L-s$ respectively. The two parts are characterized by partition
functions $Z_1 \sim \mu^s s^{\gamma_s-1}$ and $Z_2 \sim \mu^{L-s}
(L-s)^{\gamma_s-1}$, where $\gamma_s$ is an entropic exponent associated
to a polymer attached to a planar surface. One obtains the equilibrium
free energy of the polymer on the pore~\cite{chua01}:
\begin{equation}
f_t (s) = -(\gamma_s-1) \log\left[ \frac s L  
\left(1 - \frac s L\right) \right]
\label{free_transl}
\end{equation}
This free energy has a strong analogies with that of the zipping polymer
at $T_c$ (Eq.~\ref{free_tc}). One difference is the value of the exponent.
For a three dimensional self-avoiding walk attached to a planar surface
$\gamma_s= 0.70$ \cite{vand98}, whereas from renormalization group
results \cite{scha92} one finds $\sigma_1 + \sigma_3 \approx - 0.08$.

In an early study of translocation, Chuang et al.~\cite{chua01}
analyzed the FP equation corresponding to a Brownian particle moving in
the potential (\ref{free_transl}). Within this framework they showed
that, by appropriate rescaling of polymer length $L$, time and space
coordinates one finds a translocation time scaling as $\tau_t^{FP} \sim
L^2$, independently on the value of $\gamma_s$. A similar rescaling is
also applicable to the free energy in Eq.~(\ref{free_tc}), even in the
presence of an analytic scaling function $g(x)$. The analysis thus shows
that the zipping/unzipping time $\tau_{z/u}^{FP} \sim L^2$.  However it
was also argued~\cite{chua01} that this result is not self-consistent as
the time required for a polymer to equilibrate (the Rouse time, $\tau_R
\sim L^{1+2\nu}$ and $\nu \approx 0.59$ in three dimensions thus $\tau_R
\sim L^{2.18}$) turns out to be longer than the predicted $\sim L^2$
from the FP equation. In conclusion a polymer cannot translocate through
quasi-equilibrium states, because it would do so at a rate at which it
cannot equilibrate~\cite{chua01}.  A similar reasoning can be applied
to the potential of zipping strands (Eq.~(\ref{free_tc})). It can be
shown that the modulating scaling function $g(x)$ does not alter the
general result and thus the FP equation predicts a zipping time scaling
as $\tau \sim L^2$. As for translocation, however, this result is not
self-consistent since the equilibration time of the polymer would then
exceed the zipping/unzipping time. One deduces also for zipping at $T_c$
a lower bound $\tau_{z/u} \geq L^{1+2\nu}$ of the zipping/unzipping time.

\section{Simulation Results}

\subsection{Scaling of zipping/unzipping times}

The simulations at $T=T_c$ are started from a configuration with $s=s_0$,
i.e. with a star polymer with a double stranded segment of length $s_0$
and two single strands of length $s-s_0$. The critical point is obtained
by setting the Boltzmann weight of binding of two monomers to $\omega
= \mu$.  The connectivity constant for self-avoiding polymers in a
fcc lattice is known with a high degree of accuracy $\mu=10.0362(6)$
\cite{ishi89}, which provides $T_c$ with a very good precision.
The initial configuration is thermalized while keeping $s$ fixed.  This
constraint is released at time $t=0$ and the simulation is stopped once
one of the two boundaries $s=0$ or $s=L$ is reached. These boundaries
correspond to fully unzipped strands and to fully zipped strands,
respectively (see Fig.~\ref{zipunzip}(a) and (c)). The simulations are
repeated typically for $10^4$ different realizations.

\begin{figure}[t]
\centering\includegraphics[width=0.48\textwidth]{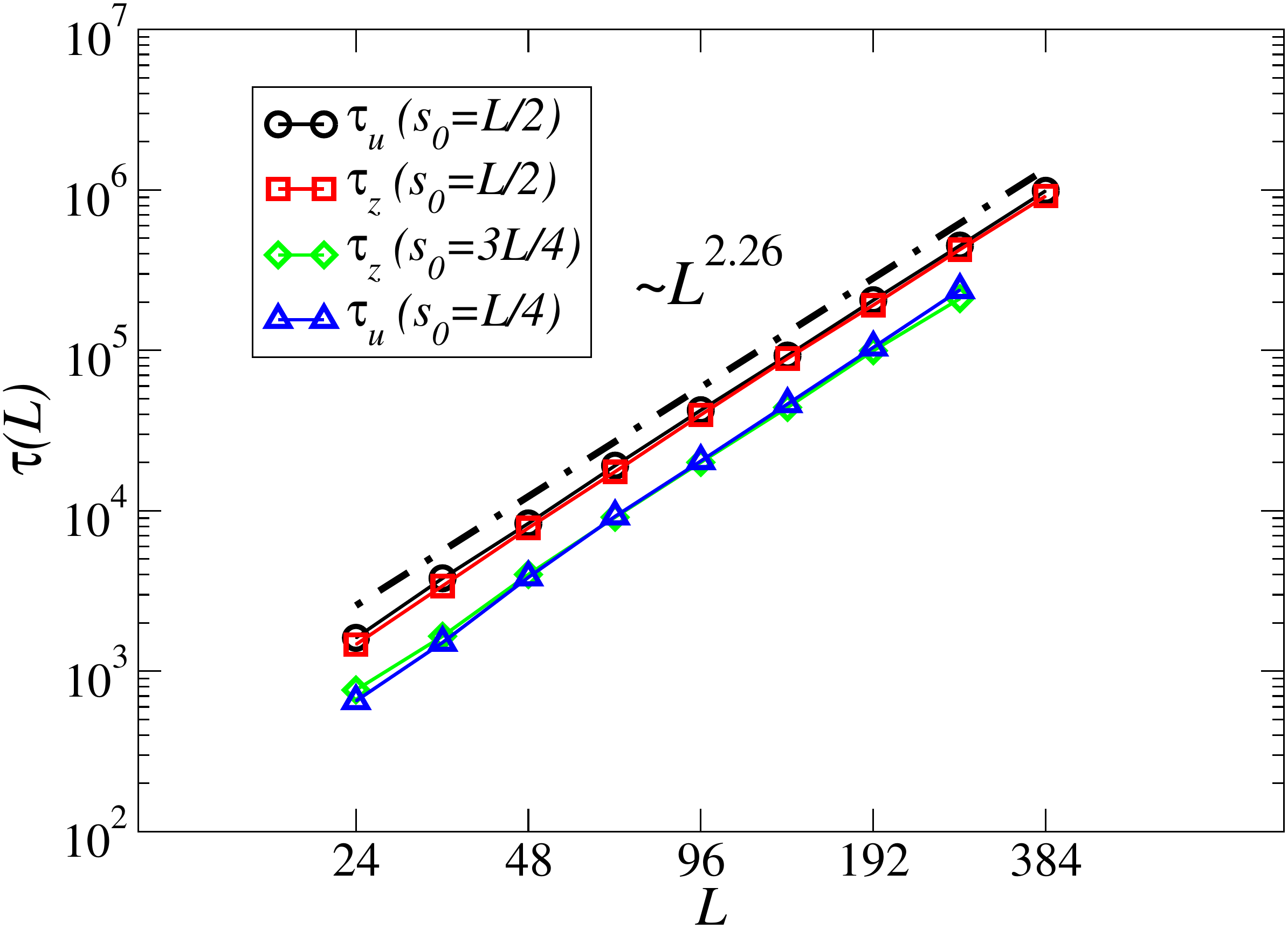}
\caption{(Color online) Log-log plot of the zipping ($\tau_{z}$) and
unzipping ($\tau_{u}$) times as functions of the strand lengths $L$.
The two quantities scale with the same anomalous exponent $\tau_{z} \sim
\tau_{u} \sim L^{2.26(2)}$. Circles and squares are data from simulations
with initial condition $s_0=L/2$, i.e. half-zipped configurations.
Diamonds correspond to the zipping time using as initial condition $s_0
= 3L/4$. Triangles correspond to the unzipping time with initial condition
$s_0 = L/4$.}
\label{tau_L}
\end{figure}

Figure~\ref{tau_L} shows a plot of the averages of the zipping $\tau_{z}$
and unzipping $\tau_{u}$ times as a function of the strand length. Error
bars are smaller than symbol sizes. The four different data sets
correspond to different initial conditions $s_0=L/2$ (half-unzipped),
$s_0 = L/4$ and $s_0 = 3L/4$. Apart from small deviations, for very short
polymers, the data show a power-law behavior which is described by the
same exponent $\tau_{z}  \sim \tau_{u} \sim L ^{2.26(2)}$. Note that for
$s_0=L/2$ the unzipping is systematically slightly slower than zipping
($\tau_{u} > \tau_{z}$). The differences are due to the asymmetry of
the problem with respect to the interchange $s \to L-s$. This asymmetry
has seemingly not a strong effect on the dynamics; it only influences
$\tau_{z}$ and $\tau_{u}$ by a multiplicative factor, but it does
not change the exponent. The exponent $\alpha = 2.26(2)$ indicates a
subdiffusive behavior. In addition the expected lower bound $\tau_R
\sim L^{1+2\nu}$, discussed above, is verified. But clearly $\alpha >
1+2\nu = 2.18$. Also notice that $\alpha$ is considerably lower than
the value $2+\nu=2.58$ recently conjectured for unbiased translocation
\cite{panj07}.

\subsection{Probability distribution function (pdf)}
 
We computed next $P(s,t)$, the probability distribution function
(pdf) of the value of $s$ at time $t$.  Initially, at time $t=0$, one
has $P(s,0)=\delta(s-s_0)$, whereafter $P(s,t)$ spreads in time. 
Here we consider the case $s_0 = L/2$. The
shape of the distribution at later times gives some insights on the
nature of the underlying stochastic process. It has been recently
suggested~\cite{zoia09,dubb11} that unbiased translocation could be
described by a class of processes known as fractional Brownian motion
(fBm). The fBm \cite{mand68} is a Gaussian, self-affine process with
stationary increments. It is described by a probability distribution with
a Gaussian shape, but with variance growing as a power law in time, i.e.
\begin{equation}
P(s,t) = \frac{1}{\sqrt{2 \pi D t^{2H}}} 
\exp\left[-\frac{(s-s_0)^2}{4Dt^{2H}}\right]
\label{fBm_pdf}
\end{equation}
where $s_0=L/2$ in our setup (the starting point of the simulations)
and $D$ is a constant. Here $0 < H < 1$ is known as the Hurst exponent;
$H=1/2$ corresponds to Brownian motion. $H < 1/2$ and $H > 1/2$ are
the subdiffusive and the superdiffusive cases.  The distribution
(\ref{fBm_pdf}) cannot hold at all times, due to the presence of
boundaries at $s=0$ and $s=L$.

\begin{figure}[t]
\centering\includegraphics[width=0.48\textwidth]{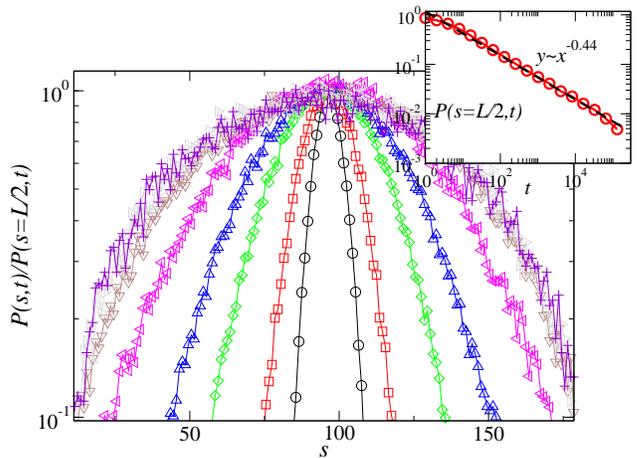}
\caption{(Color online) Plot of $P(s,t)/P(L/2,t)$ for strands of length
$L=192$ for different simulation times $t=2^9$, $2^{11}$, $2^{13}$,
$2^{14}$, $2^{15}$, $2^{16}$, $2^{17}$ and $2^{18}$. The inset shows
$P(s=L/2,t)$ versus time in a log-log plot.}
\label{pdf}
\end{figure}

Figure~\ref{pdf} (main graph) shows plots of $P(s,t)/P(L/2,t)$ vs. $s$
on a log-linear scale at different times for strands of length $L=192$,
averaged over $10^5$ histories.  In these simulations, $P(s,t)$ is
calculated from the surviving samples, i.e. those which never reach the
boundaries $s=0$ and $s=L$ at time $t$.  Similar results were obtained
for other $L$-values. One observes the spreading of the distribution
from its initial delta shape. For large times, $P(s,t)$ converges to a
stationary distribution whose properties we will discuss below.

\begin{figure}[t!]
\centering\includegraphics[width=0.48\textwidth]{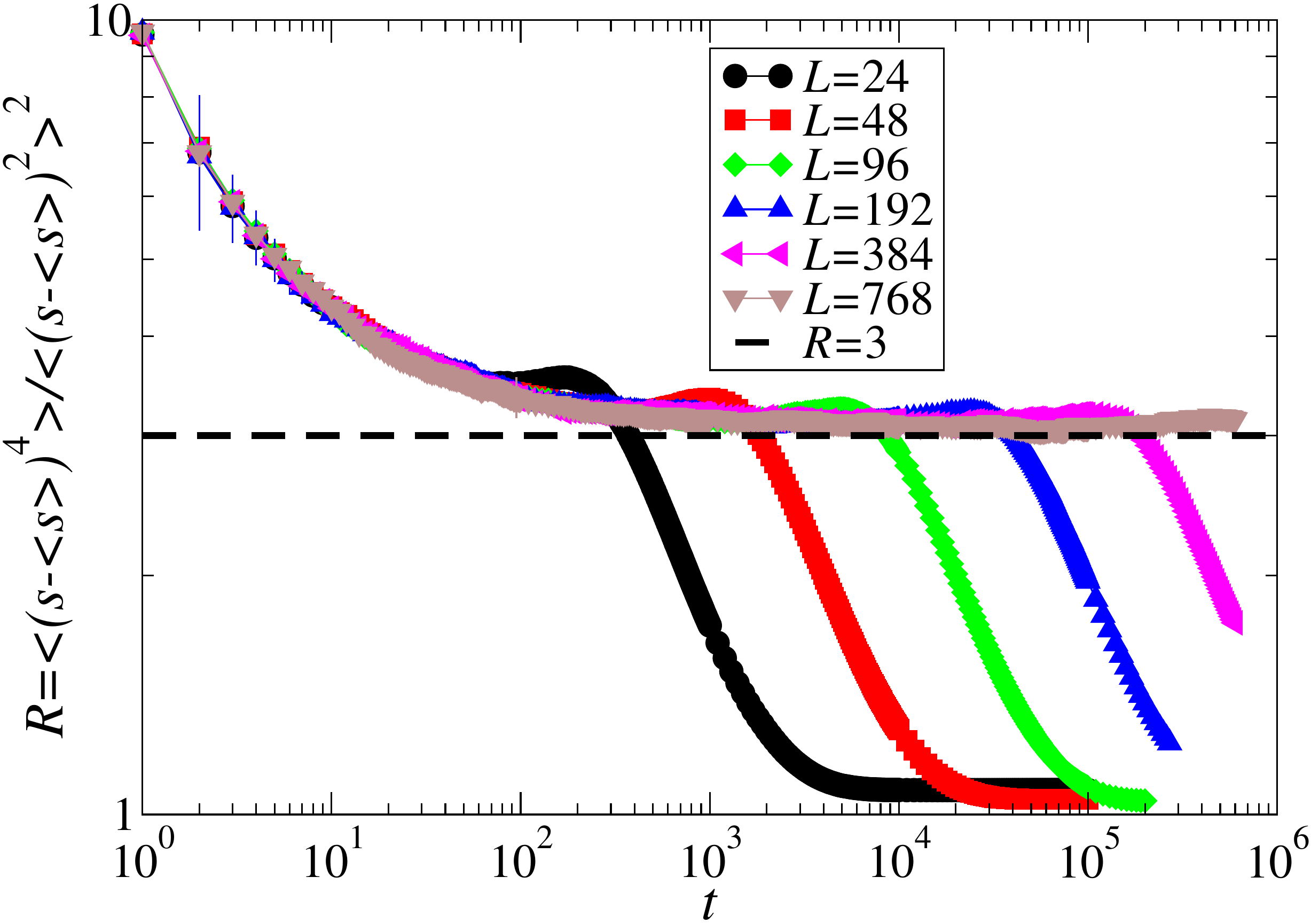}
\caption{(Color online) Plot of $R = \langle(s-\langle s\rangle)^4\rangle/
\langle(s-\langle s\rangle)^2\rangle^2$ as a function of time for
various $L$-values. The dashed line indicates the Gaussian value $R=3$.
The sampling is over $5 \cdot 10^5$ configurations for $L=24$, $48$,
$2 \cdot 10^5$ configurations for $L=96$ and $10^5$ configurations for
$L=192$, $384$, $768$.}
\label{ratio}
\end{figure}

As a check whether (and when) these data can be described by a Gaussian
distribution we have calculated the ratio $R\equiv \langle(s-\langle
s\rangle)^4\rangle /\langle(s-\langle s\rangle)^2\rangle^2$ which for
a Gaussian equals $3$. Results for $R$ as a function of time and for
various values of $L$ are shown in Figure \ref{ratio}. These data show
that after an initial regime the gaussian value is {\bf very slowly}
approached.  After a time of the order of the zipping/unzipping time,
the boundaries are being felt, and a deviation from the Gaussian value
develops, as should be expected. The extrapolation for different values
of $L$ yields $R=3.04(5)$, consistent with a gaussian distribution.

We next turn to the determination of the exponent $H$. This can be done
in two ways. Firstly, we looked at $P(s=L/2,t)$ which should decay as
$t^{-H}$. As can be seen in the inset of Fig. \ref{pdf} our results are
well fitted with a power law. From an analysis of the data for various
$L$-values we arrive at the estimate $H=0.44(1)$.

Secondly, we determined the scaling of the variance of the position
distribution,  $\sigma^2 (t) = \displaystyle\langle (s-\langle s
\rangle)^2\rangle$, as a function of time.  For the Gaussian distribution
(\ref{fBm_pdf}) one has $ \sigma^2(t) \sim t^{2H}$. For times above
$\tau_{z/u}$ deviations from this behavior appear and asymptotically in
time  $\sigma^2$ should saturate at a value $ \sim L^2$. We therefore
expect a scaling of the form
\begin{eqnarray}
\sigma^2(t) \sim t^{2H} F(t/\tau)
\label{scaling}
\end{eqnarray}
where $F(x)$ is a scaling function and $\tau$ is  $\tau_{z}$ or $\tau_{u}$.

\begin{figure}[t]
\centering\includegraphics[width=0.48\textwidth]{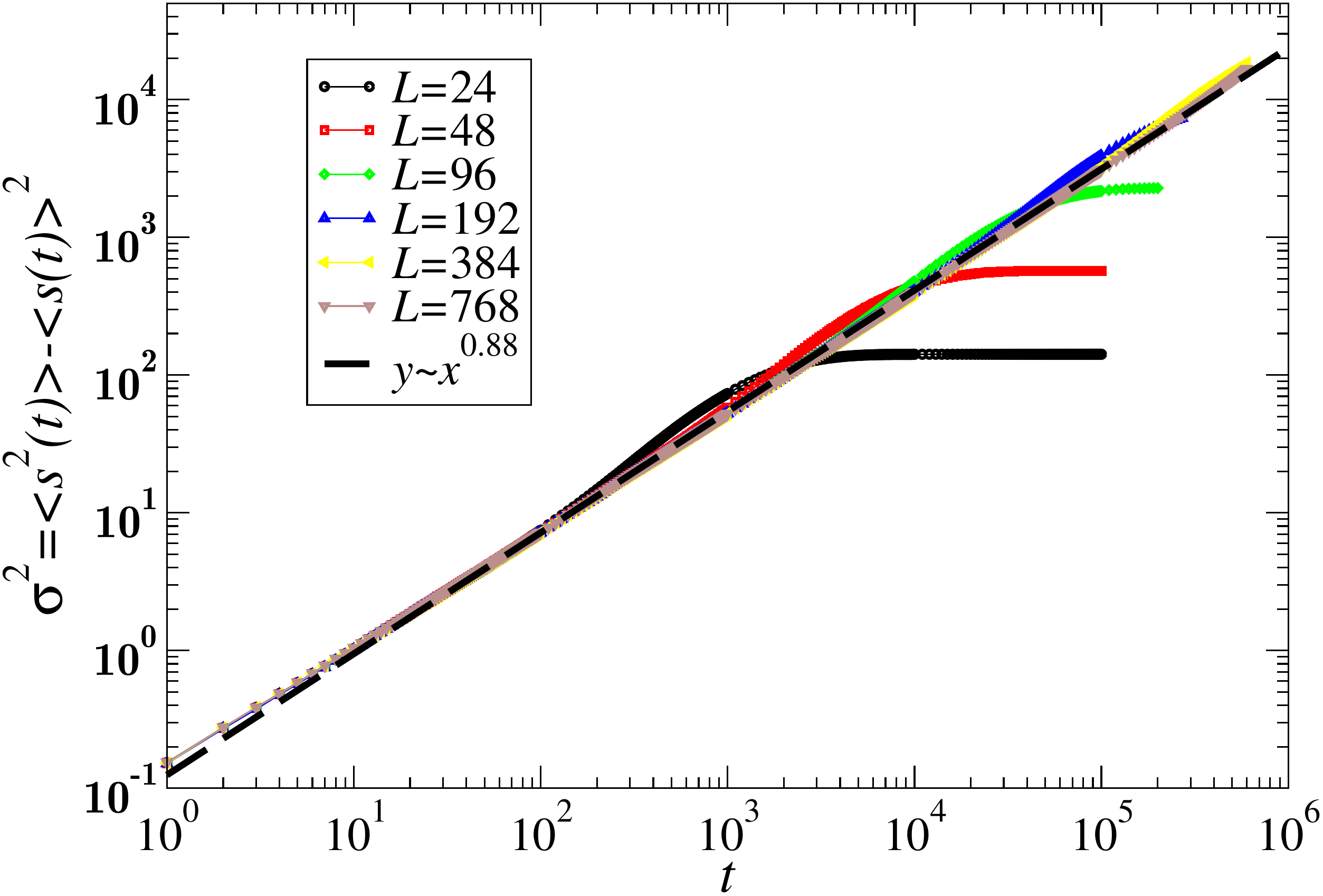}
\caption{(Color online) Plot of $\sigma^2$ as a function of time $t$
for different values of $L$. The dashed line is a power-law fit to the
data, yielding $H=0.44$.}
\label{sigma2t}
\end{figure}

This is precisely the behavior that we observe in our simulations
(Fig.~\ref{sigma2t}). From the initial power law increase we arrive at
the independent estimate of $H=0.44(1)$.  Since at $t \approx \tau,\
\sigma^2 \sim L^2$ we obtain $\tau_{z} \sim \tau_{u} \sim L^{\alpha}$
where $\alpha=1/H$. For $H=0.44(1), \alpha=2.27(2)$, fully consistent
with our results on the behavior of the zipping/unzipping times shown
in Fig.~\ref{tau_L}.  In conclusion, we find clear numerical evidence
that the dynamics of the zipping coordinate $s$ is well described by
the distribution (\ref{fBm_pdf}) with an exponent $H=0.44(1)$ in the
time regime where $t<\tau_{z/u}$.

\subsection{Velocity autocorrelation}

\begin{figure}[t]
\centering\includegraphics[width=0.48\textwidth]{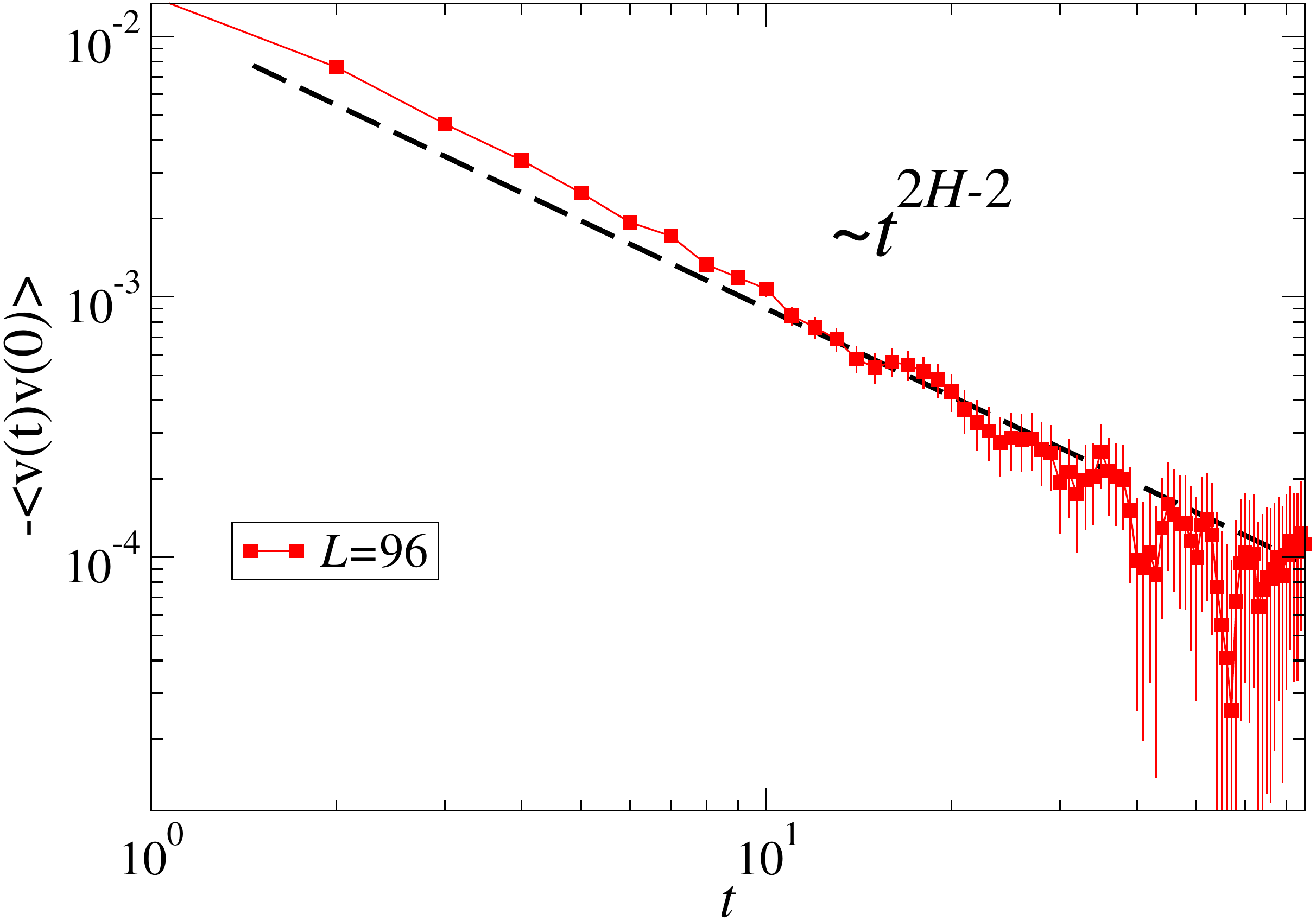}
\caption{(Color online) Log-log plot of minus the velocity autocorrelation
as a function of time. The dashed line is a straight line (not a fit)
with slope $-1.12$.}
\label{figvelcor}
\end{figure}

Since fBm  is a Gaussian process, it is fully characteristed by its
average and covariance matrix. The latter, which can also be interpreted
as a position correlation function, has the form \cite{mand68}
\begin{eqnarray}
\langle s(t_1)s(t_2)\rangle = D (t_1^{2H} + t_2^{2H} - |t_1-t_2|^{2H})
\label{poscor}
\end{eqnarray}
where $D$ is a constant. The autocorrelation of the velocity $v=ds/dt$ can 
be obtained by differentiation with respect to $t_1$ and $t_2$ \cite{Qian03}
\begin{eqnarray*}
\langle v(t_1) v(t_2)\rangle &=&\frac{d^2 \langle s(t_1) s(t_2)\rangle}{dt_1 dt_2} = -D\frac{d^2 |t_1-t_2|^{2H}}{dt_1 dt_2} \nonumber \\
&=&  2H(2H-1) D |t_1-t_2|^{2H-2}\nonumber \\ & &+ 2 H D |t_1-t_2|^{2H-1}\delta(t_1-t_2)
\end{eqnarray*}
Because of the stationarity of the process one obtains for $t>0$
\begin{eqnarray}
\langle v(t)v(0)\rangle= 2H (2H-1) D t^{2H-2}
\label{velcor}
\end{eqnarray}
For a fBm with $H<1/2$ (subdiffusive), as is the case for zipping, this
implies a negative velocity autocorrelation function. Using $H=0.44$
one finds $2H-2 = -1.12$.  In Fig. \ref{figvelcor} we show our results
for this quantity in a log-log plot. In order to compute $\langle
v(t)v(0)\rangle$ reliably more than $10^7$ configurations were sampled,
which is $100$ more than typically done for other quantities. Still, the
velocity autocorrelation could be estimated with sufficient precision only
for short times ($t<10^2$). It is therefore difficult to obtain a good
estimate of $H$ from these data. The figure shows however that $\langle
v(t) v(0)\rangle$ is negative and decays with a power law as expected. The
decay for $t>10$ is not inconsistent with the predicted value $-1.12$.

\subsection{Survival probability}

\begin{figure}[b!]
\centering\includegraphics[width=0.48\textwidth]{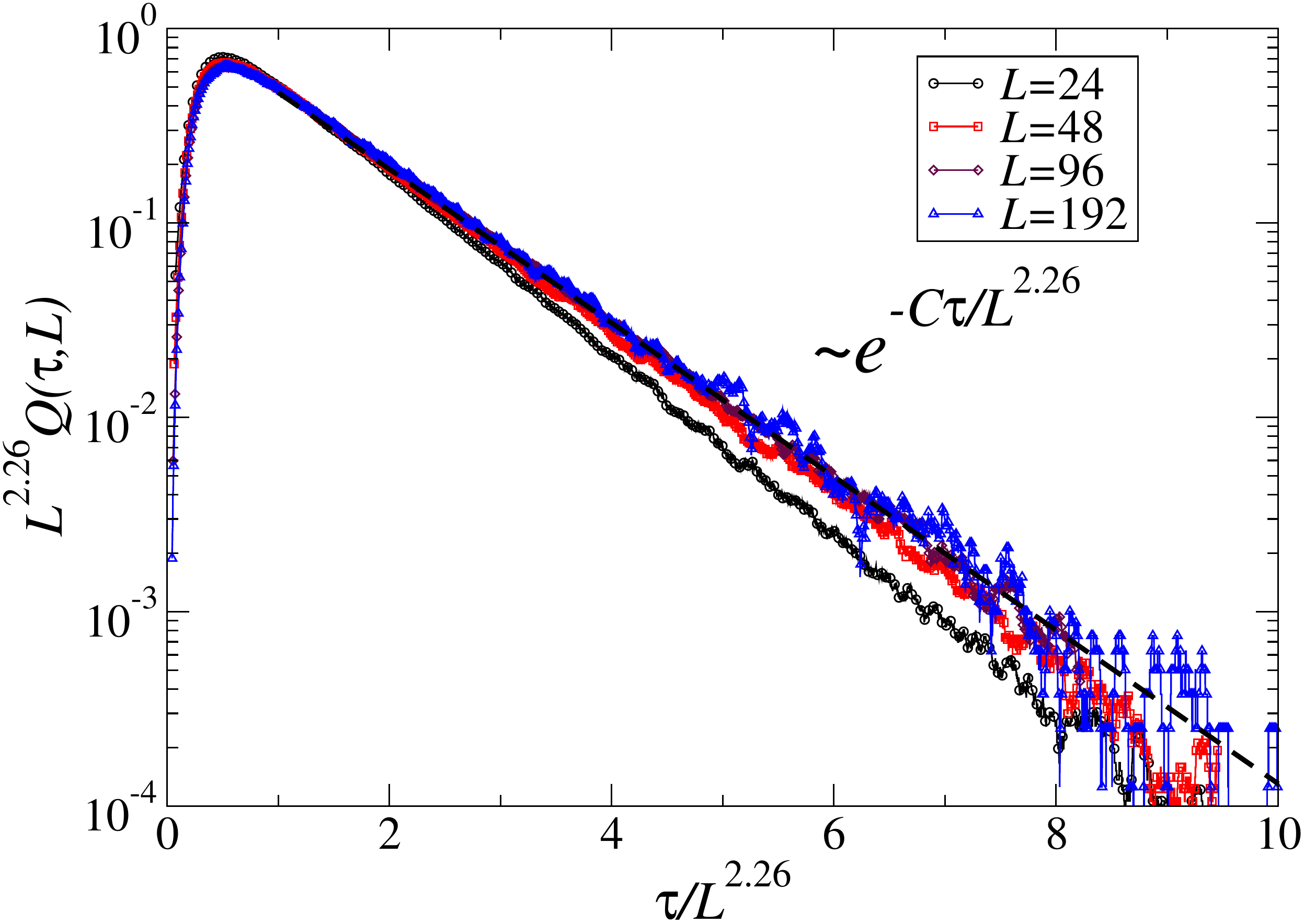}
\caption{(Color online) Linear-log plot of the scaling function of the
probability distribution of the unzipping time for the sizes $L=24$,
$48$, $96$ and $192$. The calculations are done over $10^6$ configurations
for $L=$24, 48 and 96 and over $2\cdot10^5$ configurations for $L=192$.}
\label{qt}
\end{figure}

Next, we look at the probability distribution of the (un)zipping
time $Q(\tau_{z/u},L)$. These results are plotted in Figure~\ref{qt}
for $L=24$, $48$, $96$ and $192$ (only unzipping data are shown, the
distributions for the zipping times are very similar). As seen from the
figure, the probability distribution decays exponentially for long times.
The data for different lengths collapse into a single scaling function
(for $L$ sufficiently large), where $\tau/L^{2.26}$ is used as a scaling
variable. The normalization condition of the probability distribution
function implies the following scaling form:
\begin{equation}
Q(\tau,L)=L^{-2.26}f(\tau/L^{2.26})
\end{equation}
where $f(x)$ is a scaling function which decays exponentially for large
$x$.  An exponential decay of the survival probability $S(t)$, which is
related to the pdf of the (un)zipping time by $Q(t)= - dS/dt$ is also
characteristic for fBm with two absorbing boundaries \cite{zoia09}. These
results also rule out a description of $P(s,t)$ in terms of a fractional
Fokker-Planck equation with absorbing boundaries for which $Q(\tau)$
decays as a power law \cite{yust04,gitt04}.

\subsection{Boundary exponents}

There has been quite some interest in the behavior of fractional Brownian
motion in the vicinity of an absorbing boundary. For ordinary diffusion
the probability distribution $P(s,t)$ on the interval $[0,L]$ becomes
proportional to $\sin{(\pi s/L)}$ asymptotically. This means that $P(s,t)$
vanishes linearly with the distance $y$ from an absorbing boundary.  For a
fBm with Hurst exponent $H$ it has been recently argued~\cite{zoia09} that
the pdf vanishes as $\sim y^\phi$, with $\phi = (1-H)/H$.  For ordinary
diffusion ($H=1/2$) this gives the correct result $\phi = 1$.

\begin{figure}[t]
\centering\includegraphics[width=0.48\textwidth]{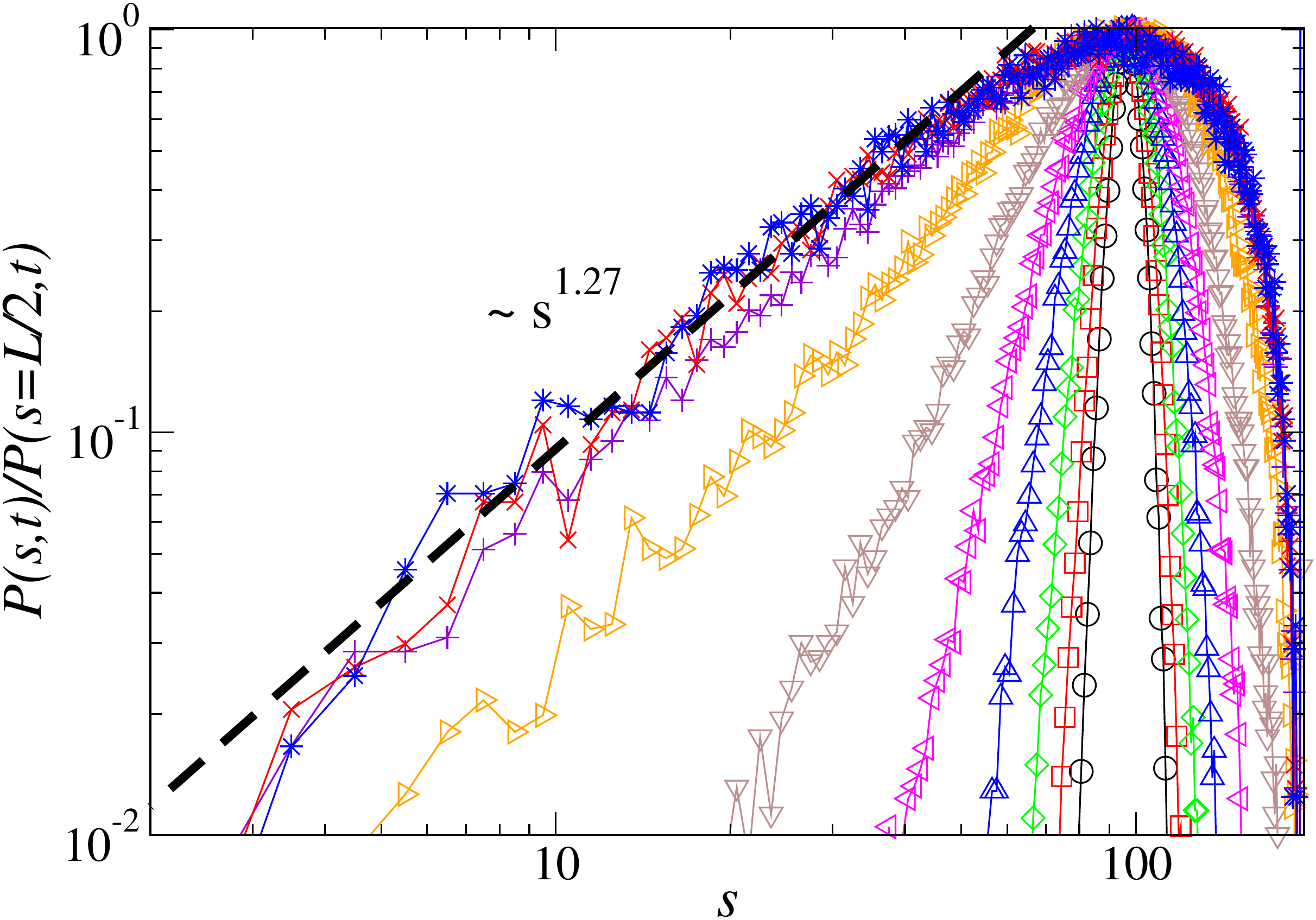}
\caption{(Color online) Plot of $\ln P(s,t)$ vs. $\ln s$ emphasizing the
vanishing of the position pdf in the vicinity of the absorbing boundaries
(same data as in Fig. \ref{pdf}, $L= 192$ and times are $t=2^9, 2^{10},
\ldots, 2^{17}, 2^{18}$).  For the two longest simulation times the
distribution is almost stationary and the probability distribution
vanishes in agreement with a power-law behavior $\sim s^{1.27}$ (dashed
line). This is the expected behavior for a fractional
Brownian motion with $H=0.44$ in the vicinity of a boundary, as derived
in Ref.~\cite{zoia09}.}
\label{boundary}
\end{figure}

In view of this recent interest, we have investigated the behavior of
the pdf near absorbing boundaries.  In order to determine the boundary
exponent $\phi$ we have replotted in Figure~\ref{boundary} the results
of Figure \ref{pdf} in a log-log plot.  According to Ref.~\cite{zoia09}
a fBm with a Hurst exponent $H = 0.44$ would imply a boundary exponent
$\phi = 1.27$. The corresponding power-law behavior is shown as a dashed
line in Figure~\ref{boundary}. This exponent is in good agreement with
the numerical data at sufficiently long simulation times for which the
distribution becomes stationary.

\section{Conclusions}

In this paper we have analyzed the dynamics of zipping polymers at their
critical point by means of simulations of polymers undergoing Rouse
dynamics.  The natural coordinate describing the process is $s$, denoting
the length of the double stranded (zipped) part of the strands. At $T=T_c$
the equilibrium free energy has a form very similar to that of a polymer
translocating through a membrane, i.e. a logarithmic dependence on $s$,
except for a weakly modulating function, which is shown to have a weak
effect on the dynamics. 

Using the same arguments as for translocation one can derive a lower bound
$\tau_{z/u} \sim L^\alpha$ with $\alpha \geq 1+2\nu \approx 2.18$. The
numerical results are indeed in agreement with this behavior, and provide
an estimate $\alpha \approx 2.26(2)$ which is 
higher than
the lower bound value. Different exponents have been reported in the
literature for polymer translocation (ranging from 2.23 \cite{wei07},
2.51\cite{chat08} to 2.59 \cite{panj07}), but it should be noticed that
critical zipping does not seem to share a common universal dynamical
scaling behavior with unforced polymer translocation. This could be a bit
surprising at first, as a recent study of zipping dynamics at $T<T_c$
showed instead a good agreement with that found in forced polymer
translocation \cite{ferr11}. On the other hand it also reported that
that the unzipping dynamics at $T>T_c$ is not anomalous ($\tau_u \sim
L$). At $T=T_c$ the dynamics is a complex combination of zipping and
unzipping resulting in an exponent that deviates from that for unbiased
translocation.

It has been recently suggested that translocation could be described
by fractional Brownian motion \cite{dubb11,zoia09,panj11}. We have therefore
considered the possibility that a fBm approach could also provide a
consistent description of the critical zipping/unzipping dynamics. Our
results show that this is indeed the case. On time scales smaller than the
zipping/unzipping time, the distribution of the zipping coordinate $s$
is well described by a Gaussian with a variance that grows as $t^{2H}$
with $H=0.44(1)$. In the presence of absorbing boundary conditions we
find that $P(s,t)$ conditionned on not being absorbed yet, converges
to a distribution that behaves as a power law near the boundaries. The
associated exponent $\phi$ agrees with that recently determined for
fBm~\cite{zoia09}.

Fractional Brownian motion has been observed in other polymer processes,
most notably in the fluctuations of the distance between an electron
transfer donor and acceptor pair in single molecule experiments
on proteins \cite{kou04}. Recently, it has been argued that this
complicated dynamics can arise from a superposition of Markovian
fluctuations of normal modes as for example those present in the Rouse
model \cite{Dua11}. It would be interesting to get further insight in the
underlying dynamics of the zipping coordinate $s$ in a similar spirit.

{\bf Acknowledgement} We would like to thank G.~Barkema, R.~Metzler,
D.~Panja and A.~Rosso for interesting discussions on the subject of
this paper.


\begin{thebibliography}{10}%
\makeatletter
\providecommand \@ifxundefined [1]{%
 \ifx #1\undefined \expandafter \@firstoftwo
 \else \expandafter \@secondoftwo
\fi
}%
\providecommand \@ifnum [1]{%
 \ifnum #1\expandafter \@firstoftwo
 \else \expandafter \@secondoftwo
\fi
}%
\providecommand \enquote [1]{``#1''}%
\providecommand \bibnamefont  [1]{#1}%
\providecommand \bibfnamefont [1]{#1}%
\providecommand \citenamefont [1]{#1}%
\providecommand\href[0]{\@sanitize\@href}%
\providecommand\@href[1]{\endgroup\@@startlink{#1}\endgroup\@@href}%
\providecommand\@@href[1]{#1\@@endlink}%
\providecommand \@sanitize [0]{\begingroup\catcode`\&12\catcode`\#12\relax}%
\@ifxundefined \pdfoutput {\@firstoftwo}{%
 \@ifnum{\z@=\pdfoutput}{\@firstoftwo}{\@secondoftwo}%
}{%
 \providecommand\@@startlink[1]{\leavevmode}%
 \providecommand\@@endlink[0]{}%
}{%
 \providecommand\@@startlink[1]{%
  \leavevmode
  \pdfstartlink
   attr{/Border[0 0 1 ]/H/I/C[0 1 1]}%
   user{/Subtype/Link/A<</Type/Action/S/URI/URI(#1)>>}%
  \relax
 }%
 \providecommand\@@endlink[0]{\pdfendlink}%
}%
\providecommand \url  [0]{\begingroup\@sanitize \@url }%
\providecommand \@url [1]{\endgroup\@href {#1}{\urlprefix}}%
\providecommand \urlprefix [0]{URL }%
\providecommand \Eprint[0]{\href }%
\@ifxundefined \urlstyle {%
  \providecommand \doi [1]{doi:\discretionary{}{}{}#1}%
}{%
  \providecommand \doi [0]{doi:\discretionary{}{}{}\begingroup
  \urlstyle{rm}\Url }%
}%
\providecommand \doibase [0]{http://dx.doi.org/}%
\providecommand \Doi[1]{\href{\doibase#1}}%
\providecommand \bibAnnote [3]{%
  \BibitemShut{#1}%
  \begin{quotation}\noindent
    \textsc{Key:}\ #2\\\textsc{Annotation:}\ #3%
  \end{quotation}%
}%
\providecommand \bibAnnoteFile [2]{%
  \IfFileExists{#2}{\bibAnnote {#1} {#2} {\input{#2}}}{}%
}%
\providecommand \typeout [0]{\immediate \write \m@ne }%
\providecommand \selectlanguage [0]{\@gobble}%
\providecommand \bibinfo [0]{\@secondoftwo}%
\providecommand \bibfield [0]{\@secondoftwo}%
\providecommand \translation [1]{[#1]}%
\providecommand \BibitemOpen[0]{}%
\providecommand \bibitemStop [0]{}%
\providecommand \bibitemNoStop [0]{.\EOS\space}%
\providecommand \EOS [0]{\spacefactor3000\relax}%
\providecommand \BibitemShut [1]{\csname bibitem#1\endcsname}%



\bibitem{lumm03}%
  \BibitemOpen
  \bibfield{author}{%
  \bibinfo {author} {\bibfnamefont{D.}~\bibnamefont{Lumma}}, \bibinfo {author}
  {\bibfnamefont{S.}~\bibnamefont{Keller}}, \bibinfo {author}
  {\bibfnamefont{T.}~\bibnamefont{Vilgis}},\ and\ \bibinfo {author}
  {\bibfnamefont{J.~O.}\ \bibnamefont{R\"adler}},\ }%
  \bibfield{journal}{%
  \Doi{10.1103/PhysRevLett.90.218301}{\bibinfo {journal} {Phys. Rev. Lett.}}\
  }%
  \textbf{\bibinfo {volume} {90}},\ \bibinfo {pages} {218301} (\bibinfo {year}
  {2003})%
  \bibAnnoteFile{NoStop}{lumm03}%



\bibitem{shus04}%
  \BibitemOpen
  \bibfield{author}{%
  \bibinfo {author} {\bibfnamefont{R.}~\bibnamefont{Shusterman}}, \bibinfo
  {author} {\bibfnamefont{S.}~\bibnamefont{Alon}}, \bibinfo {author}
  {\bibfnamefont{T.}~\bibnamefont{Gavrinyov}},\ and\ \bibinfo {author}
  {\bibfnamefont{O.}~\bibnamefont{Krichevsky}},\ }%
  \bibfield{journal}{%
  \Doi{10.1103/PhysRevLett.92.048303}{\bibinfo {journal} {Phys. Rev. Lett.}}\
  }%
  \textbf{\bibinfo {volume} {92}},\ \bibinfo {pages} {048303} (\bibinfo {year}
  {2004})%
  \bibAnnoteFile{NoStop}{shus04}%




\bibitem{petr06}%
  \BibitemOpen
  \bibfield{author}{%
  \bibinfo {author} {\bibfnamefont{E.~P.}\ \bibnamefont{Petrov}}, \bibinfo
  {author} {\bibfnamefont{T.}~\bibnamefont{Ohrt}}, \bibinfo {author}
  {\bibfnamefont{R.~G.}\ \bibnamefont{Winkler}},\ and\ \bibinfo {author}
  {\bibfnamefont{P.}~\bibnamefont{Schwille}},\ }%
  \bibfield{journal}{%
  \Doi{10.1103/PhysRevLett.97.258101}{\bibinfo {journal} {Phys. Rev. Lett.}}\
  }%
  \textbf{\bibinfo {volume} {97}},\ \bibinfo {pages} {258101} (\bibinfo {year}
  {2006})%
  \bibAnnoteFile{NoStop}{petr06}%



\bibitem{chua01}%
  \BibitemOpen
  \bibfield{author}{%
  \bibinfo {author} {\bibfnamefont{J.}~\bibnamefont{Chuang}}, \bibinfo {author}
  {\bibfnamefont{Y.}~\bibnamefont{Kantor}},\ and\ \bibinfo {author}
  {\bibfnamefont{M.}~\bibnamefont{Kardar}},\ }%
  \bibfield{journal}{%
  \bibinfo {journal} {Phys. Rev. E}\ }%
  \textbf{\bibinfo {volume} {65}},\ \bibinfo {pages} {011802} (\bibinfo {year}
  {2001}).
  \bibAnnoteFile{NoStop}{chua01}%



\bibitem{bouc90}%
  \BibitemOpen
  \bibfield{author}{%
  \bibinfo {author} {\bibfnamefont{J.-P.}\ \bibnamefont{Bouchaud}}\ and\
  \bibinfo {author} {\bibfnamefont{A.}~\bibnamefont{Georges}},\ }%
  \bibfield{journal}{%
  \Doi{DOI: 10.1016/0370-1573(90)90099-N}{\bibinfo {journal} {Physics
  Reports}}\ }%
  \textbf{\bibinfo {volume} {195}},\ \bibinfo {pages} {127 } (\bibinfo {year}
  {1990}).
  \bibAnnoteFile{NoStop}{bouc90}%



\bibitem{kou04}%
  \BibitemOpen
  \bibfield{author}{%
  \bibinfo {author} {\bibfnamefont{S.~C.}\ \bibnamefont{Kou}}\ and\ \bibinfo
  {author} {\bibfnamefont{X.~S.}\ \bibnamefont{Xie}},\ }%
  \bibfield{journal}{%
  \Doi{10.1103/PhysRevLett.93.180603}{\bibinfo {journal} {Phys. Rev. Lett.}}\
  }%
  \textbf{\bibinfo {volume} {93}},\ \bibinfo {pages} {180603} (\bibinfo {year}
  {2004})%
  \bibAnnoteFile{NoStop}{kou04}%



\bibitem{jaeh11}%
  \BibitemOpen
  \bibfield{author}{%
  \bibinfo {author} {\bibfnamefont{J.-H.}\ \bibnamefont{Jeon}}, \bibinfo
  {author} {\bibfnamefont{V.}~\bibnamefont{Tejedor}}, \bibinfo {author}
  {\bibfnamefont{S.}~\bibnamefont{Burov}}, \bibinfo {author}
  {\bibfnamefont{E.}~\bibnamefont{Barkai}}, \bibinfo {author}
  {\bibfnamefont{C.}~\bibnamefont{Selhuber-Unkel}}, \bibinfo {author}
  {\bibfnamefont{K.}~\bibnamefont{Berg-S\o{}rensen}}, \bibinfo {author}
  {\bibfnamefont{L.}~\bibnamefont{Oddershede}},\ and\ \bibinfo {author}
  {\bibfnamefont{R.}~\bibnamefont{Metzler}},\ }%
  \bibfield{journal}{%
  \Doi{10.1103/PhysRevLett.106.048103}{\bibinfo {journal} {Phys. Rev. Lett.}}\
  }%
  \textbf{\bibinfo {volume} {106}},\ \bibinfo {pages} {048103} (\bibinfo {year}
  {2011})%
  \bibAnnoteFile{NoStop}{jaeh11}%



\bibitem{panj10}%
  \BibitemOpen
  \bibfield{author}{%
  \bibinfo {author} {\bibfnamefont{D.}~\bibnamefont{Panja}},\ }%
  \bibfield{journal}{%
  \bibinfo {journal} {J. Stat. Mech.: Theory and Exp.}\ }%
  \textbf{\bibinfo {volume} {2010}},\ \bibinfo {pages} {P06011} (\bibinfo
  {year} {2010}).
  \bibAnnoteFile{NoStop}{panj10}%



\bibitem{sung96}%
  \BibitemOpen
  \bibfield{author}{%
  \bibinfo {author} {\bibfnamefont{W.}~\bibnamefont{Sung}}\ and\ \bibinfo
  {author} {\bibfnamefont{P.~J.}\ \bibnamefont{Park}},\ }%
  \bibfield{journal}{%
  \Doi{10.1103/PhysRevLett.77.783}{\bibinfo {journal} {Phys. Rev. Lett.}}\ }%
  \textbf{\bibinfo {volume} {77}},\ \bibinfo {pages} {783} (\bibinfo {year}
  {1996})%
  \bibAnnoteFile{NoStop}{sung96}%



\bibitem{muth99}%
  \BibitemOpen
  \bibfield{author}{%
  \bibinfo {author} {\bibfnamefont{M.}~\bibnamefont{Muthukumar}},\ }%
  \bibfield{journal}{%
  \bibinfo {journal} {J. Chem. Phys.}\ }%
  \textbf{\bibinfo {volume} {111}},\ \bibinfo {pages} {10371} (\bibinfo {year}
  {1999})%
  \bibAnnoteFile{NoStop}{muth99}%


\bibitem{wolt06}%
  \BibitemOpen
  \bibfield{author}{%
  \bibinfo {author} {\bibfnamefont{J.~K.}\ \bibnamefont{Wolterink}}, \bibinfo
  {author} {\bibfnamefont{G.~T.}\ \bibnamefont{Barkema}},\ and\ \bibinfo
  {author} {\bibfnamefont{D.}~\bibnamefont{Panja}},\ }%
  \bibfield{journal}{%
  \Doi{10.1103/PhysRevLett.96.208301}{\bibinfo {journal} {Phys. Rev. Lett.}}\
  }%
  \textbf{\bibinfo {volume} {96}},\ \bibinfo {pages} {208301} (\bibinfo {year}
  {2006}).
  \bibAnnoteFile{NoStop}{wolt06}%



\bibitem{panj07}%
  \BibitemOpen
  \bibfield{author}{%
  \bibinfo {author} {\bibfnamefont{D.}~\bibnamefont{Panja}}, \bibinfo {author}
  {\bibfnamefont{G.~T.}\ \bibnamefont{Barkema}},\ and\ \bibinfo {author}
  {\bibfnamefont{R.~C.}\ \bibnamefont{Ball}},\ }%
  \bibfield{journal}{%
  \bibinfo {journal} {J. Phys.: Condens. Matter}\ }%
  \textbf{\bibinfo {volume} {19}},\ \bibinfo {pages} {432202} (\bibinfo {year}
  {2007}).
  \bibAnnoteFile{NoStop}{panj07}%



\bibitem{saka10}%
  \BibitemOpen
  \bibfield{author}{%
  \bibinfo {author} {\bibfnamefont{T.}~\bibnamefont{Sakaue}},\ }%
  \bibfield{journal}{%
  \bibinfo {journal} {Phys. Rev. E}\ }%
  \textbf{\bibinfo {volume} {81}},\ \bibinfo {pages} {041808} (\bibinfo {year}
  {2010}).
  \bibAnnoteFile{NoStop}{saka10}%



\bibitem{ferr11}%
  \BibitemOpen
  \bibfield{author}{%
  \bibinfo {author} {\bibfnamefont{A.}~\bibnamefont{Ferrantini}}\ and\ \bibinfo
  {author} {\bibfnamefont{E.}~\bibnamefont{Carlon}},\ }%
  \bibfield{journal}{%
  \bibinfo {journal} {J. Stat. Mech.: Theory and Exp.}\ }%
  \textbf{\bibinfo {volume} {2011}},\ \bibinfo {pages} {P02020} (\bibinfo
  {year} {2011}).
  \bibAnnoteFile{NoStop}{ferr11}%



\bibitem{vock08}%
  \BibitemOpen
  \bibfield{author}{%
  \bibinfo {author} {\bibfnamefont{H.}~\bibnamefont{Vocks}}, \bibinfo {author}
  {\bibfnamefont{D.}~\bibnamefont{Panja}}, \bibinfo {author}
  {\bibfnamefont{G.~T.}\ \bibnamefont{Barkema}},\ and\ \bibinfo {author}
  {\bibfnamefont{R.~C.}\ \bibnamefont{Ball}},\ }%
  \bibfield{journal}{%
  \bibinfo {journal} {J. Phys.: Condens. Matter}\ }%
  \textbf{\bibinfo {volume} {20}},\ \bibinfo {pages} {095224} (\bibinfo {year}
  {2008}).
  \bibAnnoteFile{NoStop}{vock08}%




\bibitem{luo09}%
  \BibitemOpen
  \bibfield{author}{%
  \bibinfo {author} {\bibfnamefont{K.}~\bibnamefont{Luo}}, \bibinfo {author}
  {\bibfnamefont{T.}~\bibnamefont{Ala-Nissila}}, \bibinfo {author}
  {\bibfnamefont{S.-C.}\ \bibnamefont{Ying}},\ and\ \bibinfo {author}
  {\bibfnamefont{R.}~\bibnamefont{Metzler}},\ }%
  \bibfield{journal}{%
  \bibinfo {journal} {Europhys. Lett.}\ }%
  \textbf{\bibinfo {volume} {88}},\ \bibinfo {pages} {68006} (\bibinfo {year}
  {2009}).
  \bibAnnoteFile{NoStop}{luo09}%



\bibitem{mand68}%
  \BibitemOpen
  \bibfield{author}{%
  \bibinfo {author} {\bibfnamefont{B.~B.}\ \bibnamefont{Mandelbrot}}\ and\
  \bibinfo {author} {\bibfnamefont{J.~W.~V.}\ \bibnamefont{Ness}},\ }%
  \bibfield{journal}{%
  \bibinfo {journal} {SIAM Review}\ }%
  \textbf{\bibinfo {volume} {10}},\ \bibinfo {pages} {422} (\bibinfo {year}
  {1968}).
  \bibAnnoteFile{NoStop}{mand68}%




\bibitem{ferr10}%
  \BibitemOpen
  \bibfield{author}{%
  \bibinfo {author} {\bibfnamefont{A.}~\bibnamefont{Ferrantini}}, \bibinfo
  {author} {\bibfnamefont{M.}~\bibnamefont{Baiesi}},\ and\ \bibinfo {author}
  {\bibfnamefont{E.}~\bibnamefont{Carlon}},\ }%
  \bibfield{journal}{%
  \bibinfo {journal} {J. Stat. Mech.: Theory and Exp.}\ }%
  \textbf{\bibinfo {volume} {2010}},\ \bibinfo {pages} {P03017} (\bibinfo
  {year} {2010}).
  \bibAnnoteFile{NoStop}{ferr10}%



\bibitem{madr88}%
  \BibitemOpen
  \bibfield{author}{%
  \bibinfo {author} {\bibfnamefont{N.}~\bibnamefont{Madras}}\ and\ \bibinfo
  {author} {\bibfnamefont{A.~D.}\ \bibnamefont{Sokal}},\ }%
  \bibfield{journal}{%
  \bibinfo {journal} {J. Stat. Phys.}\ }%
  \textbf{\bibinfo {volume} {50}},\ \bibinfo {pages} {109} (\bibinfo {year}
  {1988}).
  \bibAnnoteFile{NoStop}{madr88}%



\bibitem{carm88}%
\BibitemOpen
  \bibfield{author}{%
  \bibinfo {author} {\bibfnamefont{I.}~\bibnamefont{Carmesin}}\ and\ \bibinfo
  {author} {\bibfnamefont{K.}\ \bibnamefont{Kremer}},\ }%
  \bibfield{journal}{%
  \bibinfo {journal} {Macromolecules}\ }%
  \textbf{\bibinfo {volume} {21}},\ \bibinfo {pages} {2819} (\bibinfo {year}
  {1988}).
  \bibAnnoteFile{NoStop}{carm88}%


\bibitem{thesis}%
\BibitemOpen
  \bibfield{author}{%
  \bibinfo {author} {\bibfnamefont{A.}~\bibnamefont{Ferrantini}},\ }%
  \emph{\bibinfo {title} {Models of polymer dynamics: DNA renaturation and zipping}}\
(\bibinfo {publisher} {PhD Thesis, KULeuven},\ \bibinfo {year} {2011})%
  \bibAnnoteFile{NoStop}{thesis}%



\bibitem{vand98}%
  \BibitemOpen
  \bibfield{author}{%
  \bibinfo {author} {\bibfnamefont{C.}~\bibnamefont{Vanderzande}},\ }%
  \emph{\bibinfo {title} {Lattice Models of Polymers}}\ (\bibinfo {publisher}
  {Cambridge University Press, Cambridge},\ \bibinfo {year} {1998})%
  \bibAnnoteFile{NoStop}{vand98}%



\bibitem{dupl86}%
  \BibitemOpen
  \bibfield{author}{%
  \bibinfo {author} {\bibfnamefont{B.}~\bibnamefont{Duplantier}},\ }%
  \bibfield{journal}{%
  \bibinfo {journal} {Phys. Rev. Lett.}\ }%
  \textbf{\bibinfo {volume} {57}},\ \bibinfo {pages} {941} (\bibinfo {year}
  {1986})%
  \bibAnnoteFile{NoStop}{dupl86}%



\bibitem{scha92}%
  \BibitemOpen
  \bibfield{author}{%
  \bibinfo {author} {\bibfnamefont{L.}~\bibnamefont{Sch{\"a}fer}}, \bibinfo
  {author} {\bibfnamefont{C.}~\bibnamefont{von Ferber}}, \bibinfo {author}
  {\bibfnamefont{U.}~\bibnamefont{Lehr}},\ and\ \bibinfo {author}
  {\bibfnamefont{B.}~\bibnamefont{Duplantier}},\ }%
  \bibfield{journal}{%
  \Doi{DOI: 10.1016/0550-3213(92)90397-T}{\bibinfo {journal} {Nuclear Physics
  B}}\ }%
  \textbf{\bibinfo {volume} {374}},\ \bibinfo {pages} {473 } (\bibinfo {year}
  {1992}).
  \bibAnnoteFile{NoStop}{scha92}%



\bibitem{ishi89}%
  \BibitemOpen
  \bibfield{author}{%
  \bibinfo {author} {\bibfnamefont{T.}~\bibnamefont{Ishinabe}},\ }%
  \bibfield{journal}{%
  \Doi{10.1103/PhysRevB.39.9486}{\bibinfo {journal} {Phys. Rev. B}}\ }%
  \textbf{\bibinfo {volume} {39}},\ \bibinfo {pages} {9486} (\bibinfo {year}
  {1989})%
  \bibAnnoteFile{NoStop}{ishi89}%

\bibitem{zoia09}%
  \BibitemOpen
  \bibfield{author}{%
  \bibinfo {author} {\bibfnamefont{A.}~\bibnamefont{Zoia}}, \bibinfo {author}
  {\bibfnamefont{A.}~\bibnamefont{Rosso}},\ and\ \bibinfo {author}
  {\bibfnamefont{S.~N.}\ \bibnamefont{Majumdar}},\ }%
  \bibfield{journal}{%
  \bibinfo {journal} {Phys. Rev. Lett.}\ }%
  \textbf{\bibinfo {volume} {102}},\ \bibinfo {pages} {120602} (\bibinfo {year}
  {2009})%
  \bibAnnoteFile{NoStop}{zoia09}%

\bibitem{dubb11}%
  \BibitemOpen
  \bibfield{author}{%
  \bibinfo {author} {\bibfnamefont{J.~L.~A.}\ \bibnamefont{Dubbeldam}},
  \bibinfo {author} {\bibfnamefont{V.~G.}\ \bibnamefont{Rostiashvili}},
  \bibinfo {author} {\bibfnamefont{A.}~\bibnamefont{Milchev}},\ and\ \bibinfo
  {author} {\bibfnamefont{T.~A.}\ \bibnamefont{Vilgis}},\ }%
  \bibfield{journal}{%
  \Doi{10.1103/PhysRevE.83.011802}{\bibinfo {journal} {Phys. Rev. E}}\ }%
  \textbf{\bibinfo {volume} {83}},\ \bibinfo {pages} {011802} (\bibinfo {year}
  {2011})%
  \bibAnnoteFile{NoStop}{dubb11}%


\bibitem{Qian03}
H. Qian, in {\it Processes with Long-Range Correlations: Theory and
Applications}, Lecture Notes in Physics, Vol. 621, edited by G. Rangarajan
and M.Z. Ding (Springer, New York, 2003), p. 22.


\bibitem{yust04}%
  \BibitemOpen
  \bibfield{author}{%
  \bibinfo {author} {\bibfnamefont{S.~B.}\ \bibnamefont{Yuste}}\ and\ \bibinfo
  {author} {\bibfnamefont{K.}~\bibnamefont{Lindenberg}},\ }%
  \bibfield{journal}{%
  \bibinfo {journal} {Phys. Rev. E}\ }%
  \textbf{\bibinfo {volume} {69}},\ \bibinfo {pages} {033101} (\bibinfo {year}
  {2004})%
  \bibAnnoteFile{NoStop}{yust04}%


\bibitem{gitt04}%
  \BibitemOpen
  \bibfield{author}{%
  \bibinfo {author} {\bibfnamefont{M.}~\bibnamefont{Gitterman}},\ }%
  \bibfield{journal}{%
  \bibinfo {journal} {Phys. Rev. E}\ }%
  \textbf{\bibinfo {volume} {69}},\ \bibinfo {pages} {033102} (\bibinfo {year}
  {2004})%
  \bibAnnoteFile{NoStop}{gitt04}%


\bibitem{wei07}%
  \BibitemOpen
  \bibfield{author}{%
  \bibinfo {author} {\bibfnamefont{D.}~\bibnamefont{Wei}}, \bibinfo {author}
  {\bibfnamefont{W.}~\bibnamefont{Yang}}, \bibinfo {author}
  {\bibfnamefont{X.}~\bibnamefont{Jin}},\ and\ \bibinfo {author}
  {\bibfnamefont{Q.}~\bibnamefont{Liao}},\ }%
  \bibfield{journal}{%
  \Doi{10.1063/1.2735627}{\bibinfo {journal} {J. Chem. Phys.}}\ }%
  \textbf{\bibinfo {volume} {126}},\ \bibinfo {pages} {204901} (\bibinfo {year}
  {2007})%
  \bibAnnoteFile{NoStop}{wei07}%


\bibitem{chat08}%
  \BibitemOpen
  \bibfield{author}{%
  \bibinfo {author} {\bibfnamefont{C.}~\bibnamefont{Chatelain}}, \bibinfo
  {author} {\bibfnamefont{Y.}~\bibnamefont{Kantor}},\ and\ \bibinfo {author}
  {\bibfnamefont{M.}~\bibnamefont{Kardar}},\ }%
  \bibfield{journal}{%
  \bibinfo {journal} {Phys. Rev. E}\ }%
  \textbf{\bibinfo {volume} {78}},\ \bibinfo {pages} {021129} (\bibinfo {year}
  {2008})%
  \bibAnnoteFile{NoStop}{chat08}%


\bibitem{panj11}%
  \BibitemOpen
  \bibfield{author}{%
  \bibinfo {author} {\bibfnamefont{D.}\ \bibnamefont{Panja}},\ }%
  \bibfield{journal}{%
  {\bibinfo {journal} {J. Phys. : Condens. Matter}}\ }%
  \textbf{\bibinfo {volume} {23}},\ \bibinfo {pages} {105103} (\bibinfo {year}
  {2011})%
  \bibAnnoteFile{NoStop}{panj11}%


\bibitem{Dua11}
A. Dua and R. Adhikari, J. Stat. Mech .: Theory and Exp., P04017 (2011).

\end{thebibliography}

%
\end{document}